%% file: main.tex
\documentclass[conference, a4paper]{IEEEtran}
\IEEEoverridecommandlockouts
\usepackage{cite}
\usepackage{amsmath,amssymb,amsfonts,bm}
\usepackage{mathrsfs}
\usepackage{algorithm}
\usepackage[noend]{algpseudocode}
\usepackage{graphicx}
\usepackage{comment}
\usepackage{hyperref}
\usepackage{textcomp}
\usepackage{xcolor}
\usepackage{array}
\usepackage{multirow}
\definecolor{niceblue}{rgb}{0, 0.5, 1.0}
\hypersetup{
    colorlinks=true,
    linkcolor=niceblue,
    filecolor=magenta,      
    urlcolor=niceblue,
    citecolor=niceblue,
    }

\def\BibTeX{{\rm B\kern-.05em{\sc i\kern-.025em b}\kern-.08em
    T\kern-.1667em\lower.7ex\hbox{E}\kern-.125emX}}

\begin{document}

\title{
Scalable Bilevel Optimization for Generating Maximally Representative OPF Datasets \\
\thanks{This work is supported in part by the HORIZON-MSCA-2021 Postdoctoral Fellowship Program, Project \#101066991 -- TRUST-ML. Ignasi Ventura Nadal was supported by the ERC Starting Grant VeriPhIED, funded by the European Research Council, Grant Agreement 949899.}
}

\author{\IEEEauthorblockN{Ignasi Ventura Nadal}
\IEEEauthorblockA{\textit{Department of Wind and Energy Systems} \\
\textit{Technical University of Denmark}\\
Kongens Lyngby, Denmark \\
ignad@dtu.dk}
\and
\IEEEauthorblockN{Samuel Chevalier}
\IEEEauthorblockA{\textit{Department of Wind and Energy Systems} \\
\textit{Technical University of Denmark}\\
Kongens Lyngby, Denmark \\
schev@dtu.dk}
}
\IEEEoverridecommandlockouts
\IEEEpubid{\makebox[\columnwidth]{979-8-3503-9678-2/23/\$31.00~\copyright2023 IEEE \hfill}\hspace{\columnsep}\makebox[\columnwidth]{ }}
\maketitle
\IEEEpubidadjcol
\begin{abstract}
New generations of power systems, containing high shares of renewable energy resources, require improved data-driven tools which can swiftly adapt to changes in system operation. Many of these tools, such as ones using machine learning, rely on high-quality training datasets to construct probabilistic models. Such models should be able to accurately represent the system when operating at its limits (i.e., operating with a high degree of ``active constraints"). However, generating training datasets that accurately represent the many possible combinations of these active constraints is a particularly challenging task, especially within the realm of nonlinear AC Optimal Power Flow (OPF), since most active constraints cannot be enforced explicitly. Using bilevel optimization, this paper introduces a data collection routine that sequentially solves for OPF solutions which are ``optimally far" from previously acquired voltage, power, and load profile data points. The routine, termed RAMBO, samples critical data close to a system's boundaries much more effectively than a random sampling benchmark. Simulated test results are collected on the 30-, 57-, and 118-bus PGLib test cases.
\end{abstract}


\section{Introduction}
\input{sec1}

\section{Methodology}
\input{sec2}

\section{Assessing the quality of the datasets}
\input{sec3}

\section{Numerical study}
\input{sec4}

\section{Conclusions}
\input{sec5}



\section*{Appendix}
\subsection{Application of KKT Conditions to the OPF Problem}


The OPF formulation \eqref{eqs: finalopf} considers 5 equality constraints and 9 inequality constraints, all seen in \eqref{eq: OPF_constraints} and \eqref{eq: inj}. The equality constraints are the angle reference, the power balances and the power flows. The inequality constraints are generation limits, phase difference limits, line flow limits, and voltage limits. Each equality, and inequality equation is assigned a dual variable $\lambda$, $\mu$, respectively.

These equations and duals are used to build the complementary slackness, dual feasibility and stationary conditions. The number of stationary conditions depends on the number of decision variables in the problem and contains the partial derivatives of the Lagrangian function with respect to each decision variable $x \triangleq \{v,\theta,p_G,q_G, p^{({\rm ft})}, q^{({\rm ft})}, p^{({\rm tf})}, q^{({\rm tf})}  \}$. Therefore, there are eight partial derivatives of the Lagrangian function, which consists of the OPF cost function and the primal constraints multiplied by their respective dual variables. The stationarity conditions are then given by
\begin{align*} \label{lagrangianderiv}
\frac{\partial\mathcal{L}}{\partial x}=\left[\frac{\partial\mathcal{L}}{\partial v}^{T}\frac{\partial\mathcal{L}}{\partial\theta}^{T}\frac{\partial\mathcal{L}}{\partial p_{G}}^{T}\frac{\partial\mathcal{L}}{\partial q_{G}}^{T}\frac{\partial\mathcal{L}}{\partial p^{({\rm ft/tf})}}^{T}\frac{\partial\mathcal{L}}{\partial q^{({\rm ft/tf})}}^{T}\right]^{T}\!\!\triangleq0.
\end{align*}
\bibliographystyle{IEEEtran}
\bibliography{main.bib}

\end{document}

%% file: sec1.tex
There is a growing need for power system operators and researchers to gather high-quality datasets of Optimal Power Flow (OPF) solutions. Many current OPF data collection methods are based on random sampling and only depict small regions of the feasible space. Having accurate representations of the space of OPF solutions can pave the way for building models which improve operational decision-making and enhance network performance, especially in renewable-dominated systems. Better data collection routines can be beneficial in several other areas, including (i) planning and operation of power systems, (ii) validation of simulation models, (iii) identification of system constraints, and (iv) load flow study research and development. The need for high-quality datasets has also increased due to the booming of machine learning (ML). ML-based methods are particularly sensitive to the training data \cite{stiasny2022closing}. Recent publications in the OPF literature~\cite{rahulsnew} show that adjusting training data can improve worst-case performance guarantees by up to 80\%.

Until recently, OPF research has not given much attention to dataset generation. Datasets have been generally seen as a necessary requirement for developing and demonstrating other important techniques and applications. Therefore, methods for dataset generation have often been overlooked. However, 
as the importance of training dataset \textit{quality} has come into focus, more relevance has been given to the proper training data collection. Key papers that learn OPF models initially used uniform distributions to sample the feasible space \cite{simplesample, simplesample1, simplesample2, simplesample3, simplesample5}. Authors in \cite{simplesample4} employed a Latin hypercube to generate samples between 60\% and 100\% of the nominal loads. In \cite{gaussian}, the authors utilized a truncated Gaussian distribution to generate the training samples. Other common sampling approaches used historical data \cite{historicaldata, historicaldata1}. Many of these approaches can sample load profiles very efficiently, but they cannot explicitly capture the intricate relation that loading has with generation, voltage and flows. Thus, the sampling routines are agnostic to active constraint enforcement and have difficulty capturing data on power system constraint set edges. Developing reliable models from these datasets is exceptionally challenging, as system limits aren't well-represented.

A new wave of research is tackling the problem of ``constraint aware" sampling in various ways. The OPF-Learn tool~\cite{Jones:2022} used hit-and-run sampling and infeasibility certificates to generate data with a very high number of unique active sets. Split-based sequential sampling~\cite{Strbac:2023} chooses samples to optimize a volume-based coverage metric. Authors in \cite{Rossi:2022} used Latin hypercube sampling biased towards high information content regions. Similarly, \cite{Thams:2020} utilized a directed walk (i.e.,  gradient descent) to iteratively push operating points closer to a stability boundary. More explicitly, \cite{Hu:2023} designed a learning-based model which directly outputs its own constraint violation, thus allowing for
``worth-learning" data collection.


This paper offers a new perspective on OPF dataset generation: we pose a bilevel optimization problem whose goal is to find a set of \textit{optimal} power flow solutions (lower-level) which are \textit{maximally} far apart (upper-level).
The proposed bilevel optimization routine aims to capture a significantly broader scope of OPF solution space data by implicitly sampling from its edges. The dataset's increased quality and range can lead to data-driven models that are better at understanding the physical limitations of the system. 

Our proposed data collection routine has been inspired by previous work in \cite{9960969}; this work explicitly investigated various objective functions for maximizing the distance between feasible power flow solutions. The best-performing objective functions used ``log of $\ell_1$" distance metrics across previously identified power and voltage profiles to find remote areas in the feasible space. In this present work, we utilize a distance metric function, along with a new set of optimality constraints, in order to optimally explore the OPF solution space (i.e., the set of OPF solutions for an admissible loading range).


To the best of our knowledge, this work is the first to introduce an AC-OPF sampling routine that explicitly selects OPF load inputs to optimize the distance between the voltage and generation solution profiles. This approach creates a representation of the OPF solution space which is notably better than a random sampling-base benchmark. Following are our primary contributions:



\begin{enumerate}
    \item We introduce a bilevel optimization routine for generating OPF solutions which are maximally far apart.
    \item We introduce a series of reformulations which help make the data collection routine more tractable, more accurate, and more robust to divergence.
    \item We post the data collection routine in a public repository~\cite{github_RAMBO} so that other researchers can utilize this tool.
\end{enumerate}

%% file: sec2.tex
This section introduces the OPF formulation and how it is applied in bilevel optimization to create high-quality datasets.

\subsection{OPF formulation}
The OPF problem is one of the most important problems in power systems optimization. Power systems are governed by many physical conditions and operational parameters that make it difficult for operators to determine the ideal settings. OPF aims to determine these best-operating conditions by encoding the operation in a set of equality and inequality constraints. 
In this paper, we leverage the formulation from \cite{5971792} which complies with the PowerModels framework \cite{8442948}:
\begin{subequations}\label{eq: OPF_constraints}
\begin{align}
v_{i}^{{\rm min}} & \leq v_{i}\leq v_{i}^{{\rm max}}, &  & \forall i\in\mathcal{N}\\
p_{i} & =p_{G,i}-p_{D,i}, &  & \forall i\in\mathcal{N}\\
q_{i} & =q_{G,i}-q_{D,i} &  & \forall i\in\mathcal{N}\\
p_{G,j}^{{\rm min}} & \leq p_{G,j}\leq p_{G,j}^{{\rm max}}, &  & \forall j\in\mathcal{G}\\
q_{G,j}^{{\rm min}} & \leq q_{G,j}\leq q_{G,j}^{{\rm max}}, &  & \forall j\in\mathcal{G}\\
\left|\theta_{i}-\theta_{j}\right| & \le\theta_{ij}^{{\rm max}}, &  & \forall\{i,j\}\in\mathcal{L}\\
|p_{ij}+jq_{ij}| & \le s_{ij}^{{\rm max}}, &  & \forall\{i,j\}\in\mathcal{L}\\
|p_{ji}+jq_{ji}| & \le s_{ji}^{{\rm max}}, &  & \forall\{i,j\}\in\mathcal{L}\\
\theta_{i} & =0, &  & i=1,
\end{align}
\end{subequations}
where ${\mathcal N} \triangleq \{1,2,...,n\}$ denotes the set of buses, ${\mathcal G}\subset {\mathcal N}$ denotes the subset of dispatchable generators , and  ${\mathcal L}\subset {\mathcal N}\times {\mathcal N}$ denotes the set of lines. Nodal voltage phasors are expressed by ${\bm v}e^{j{\bm \theta}}\in {\mathbb C}^{n\times1}$, and complex nodal power injections are indicated by ${\bm p}+j{\bm q}\in {\mathbb C}^{n\times1}$.

The aforementioned constraint model is linked through the nonlinear power flow equations described in (\ref{eq: inj}):
\begin{subequations}\label{eq: inj}
\begin{align}\label{p_{i}nj}p_{i} & =v_{i}\sum_{k\in\mathcal{K}_{i}}v_{k}\left(G_{ik}\cos(\theta_{ik})+B_{ik}\sin(\theta_{ik})\right),\forall i\in\mathcal{N}\\
q_{i} & =v_{i}\sum_{k\in\mathcal{K}_{i}}v_{k}\left(G_{ik}\sin(\theta_{ik})-B_{ik}\cos(\theta_{ik})\right),\forall i\in\mathcal{N}\label{q_{i}nj}
\end{align}
\end{subequations}
where ${\mathcal K}_i$ is the set of buses tied to the bus $i$. Notably, \eqref{eq: inj} also codifies $p_{ij}$, $q_{ij}$ in (\ref{eq: OPF_constraints}) as the active and reactive line flow equations for a given set of buses $\{i,j\}$.

This power system constraint model can be optimized considering a wide range of objective functions. For the sake of demonstrating the OPF data collection routine, a quadratic generation cost function is used, resulting in
\begin{subequations} \label{eqs: finalopf}
\begin{align}{\rm OPF}(p_d,q_d)\, \triangleq \, \min_{p,q,v} & \quad p_{G}^{T}C_{q}p_{G}+c_{l}^{T}p_{G}\\
\quad{\rm s.t.} & \quad\eqref{eq: OPF_constraints}-\eqref{eq: inj},
\end{align}
\end{subequations}
where $c_l^{T}$ ($C_q^{T}$) is a vector (diagonal matrix) of linear (quadratic) costs associated with each term of the vector $p_G$. This formulation is developed in a per-unit system and does not include high-voltage direct current (HVDC) links.

\subsection{Introduction to bilevel optimization reformulation}\label{subsec: intro_bilevel}

A generic bilevel optimization problem is stated in \eqref{eq: generic}. It consists of two levels of interdependent optimization problems: an upper-level and a lower-level. 
The lower-level represents a parametric optimization problem which is treated as a constraint in the upper-level problem. 
Therefore, the solution needs to fulfil optimality for both problems; otherwise, there is no feasible solution \cite{bileveltheory}:
\begin{subequations}\label{eq: generic}
\begin{alignat}{3}
\max_{x,y}\quad & f(x,y) & \\
{\rm s.t.}\quad & r(x,y)\le 0\label{eq: rxy}\\
& \min_{y} &&\mkern-35mu c(x,y)\\
& \;\;{\rm s.t.} &&\mkern-35mu g(x,y)\le0\label{eq: gxy}\\
&  && \mkern-35mu h(x,y)=0\label{eq: hxy}.
\end{alignat}
\end{subequations}
In this formulation, $r(\cdot)$ represents upper-level range constraints, while $g(\cdot)$ and $h(\cdot)$ represent lower-level inequality and equality constraints, respectively.


To solve this problem, we assume that the lower-level problem is sufficiently regular (but not necessarily convex). We explicitly reformulate the lower-level problem using the Karush–Kuhn–Tucker (KKT) conditions\cite{bilevelapp,Buason:2022}. In doing so, the problem is reduced to a single-level optimization problem with optimality conditions encoded as constraints; these include stationarity, primal/dual feasibility, and complementary constraints. The reformulation is given by
\begin{subequations}\label{eq: explicit}
\begin{align}\max_{x,y,\mu,\lambda}\quad & f(x,y)\\
\text{s.t.}\quad & \eqref{eq: rxy},\eqref{eq: gxy},\eqref{eq: hxy}\\
 & \mu\ge0\\
 & \mu\cdot g(x,y)=0\\
 & \nabla_{y}\mathcal{L}(x,y,\mu,\lambda)=0,
\end{align}
\end{subequations}
where $\mu$ and $\lambda$ are dual variables, and where the Lagrangian function is given by
\begin{align}\label{eq: lag}
\mathcal{L}(x,y,\mu,\lambda)\triangleq c(x,y)+\mu^{T}g(x,y)+\lambda^{T}h(x,y).
\end{align}
Due to the irregular complementary slackness condition, reformulation \eqref{eq: explicit} is a challenging problem for existing optimizers. 
Therefore, as proposed in previous works~\cite{bilevelapp}, we relax the complementary slackness as follows:
\begin{equation}\label{eq: relax_eps}
    \mu_j g_j(x,y) \ge - \epsilon, \quad \forall j\\ 
\end{equation}
where $\epsilon$ is sufficiently small such that the error it introduces is typically negligible. The benefit of this relaxation, however, is a significant increase in solution speed and a lower divergence rate of the interior point method used to solve this problem. 




\subsection{Bilevel optimization for OPF data collection}
The bilevel formulation \eqref{eq: explicit} is now used to create an OPF data sampling routine based on the ideas described in our previous work \cite{9960969}. In this previous work, we defined a routine to sample feasible (not optimal) power flow data, where each new sequential solution was selected to be maximally far from all previously found solutions in some library ${\mathcal S}$.


In this work, we additionally enforce optimality conditions when finding new power flow solutions. The resulting OPF data collection routine is defined in \eqref{eq: first imp}, where the upper-level constraints bound the range that OPF loads can take, while the lower-level problems solve the OPF problem via \eqref{eqs: finalopf}:
\begin{subequations} \label{eq: first imp}
\begin{align}
f_{\rm samp}(\mathcal S) \, \triangleq \, {\max_{x,p_d,q_d}}\quad & f_{{\rm dist}}(p_G,p_d,v,\mathcal{S})\\ \label{firstcons}
{\rm s.t.}\quad & p_d^{\rm min} \leq p_d \leq p_d^{\rm max} \\ \label{secondcons}
& q_d^{\rm min} \leq q_d \leq q_d^{\rm max} \\ \label{thirdcons}
& (p_G,v)\leftarrow{\rm OPF}(p_d,q_d),
\end{align}
\end{subequations}
where $x$ represents the full set of typical OPF system states:
\begin{align}
x=\{p_{G},q_{G},v,\theta,p^{{\rm ft/tf}},q^{{\rm ft/tf}}\},
\end{align}
and where we use $(p_G,v)\leftarrow{\rm OPF}(p_d,q_d)$ in the lower-level as a shorthand function which maps given load profiles to the optimal power and voltage dispatch points in the network (the associated objective and constraints are defined in \eqref{eqs: finalopf}). The upper-level objective function $f_{{\rm dist}}(p_G,p_d,v,\mathcal{S})$ is based on the best-performing functions identified by \cite{9960969}, given here as
\begin{align}
f_{{\rm dist}}= & \sum_{j\in\Gamma_\mathcal{S}}\Bigg(\log\bigg(\sum_{i\in{\mathcal{G}}}\left|p_{G_{i}}-P_{G_{i},j}\right|\bigg)+\nonumber\\
 \log\bigg(&\sum_{i\in{\mathcal{D}}}\left|p_{d_{i}}-P_{d_{i},j}\right|\bigg)+\log\bigg(\sum_{k\in{\mathcal{N}}}\left|v_{k}-V_{k,j}\right|\bigg)\!\!\Bigg),\label{upperlevelobj}
\end{align}
where $\Gamma_\mathcal{S}$ is the set of all previously sampled solutions in the data collection routine, $\mathcal{S}$ represents the library of such points, $p_{G_i}$, $p_{d_i}$, $v_k$ are decision variables, and $P_{G_i,j}$, $P_{d_i,j}$, $V_{k,j}$ are numerical values from the previously sampled solution library $\mathcal{S}=\{\cdots P_{G,j},P_{d,j},V_{j}\cdots\}$. 
The number of decision variables considered in the objective function can  potentially be increased to improve the objective's performance. However, the addition of terms comes with an increase in both problem complexity and solution time. Thus, only voltage magnitude, load power, and generation power are considered.

Our overall data collection routine is explained in Algorithm \ref{algo:iterative}. The first step is to compute a random OPF solution ${\bm s}_1^{\star}$ that will serve as the starting point of the algorithm; this first solution initializes the OPF solutions library ${\mathcal S}$. 
The routine then computes new OPF solutions that are maximally ``far away" from the previously saved solutions. This process is repeated iteratively $N$ times, with the complexity of the objective function increasing as the iteration number increases. 

Notably, in solving \eqref{eq: first imp}, we reformulate the lower-level OPF function \eqref{thirdcons} using the KKT reformulation introduced in subsection \ref{subsec: intro_bilevel} along with the relaxation in \eqref{eq: relax_eps}. The interested reader may refer to the Appendix for details.
\begin{algorithm}
\caption{OPF Data Collection Routine}\label{algo:iterative}

{\small \textbf{Require:}
OPF model, number of data points $N$, $\epsilon$

\begin{algorithmic}[1]

\State Compute random feasible solution ${\bm s}^{\star}_1$ for \eqref{eqs: finalopf}

\State Set ${\mathcal S} = {\bm s}^{\star}_1$ and $i=2$

\For{$i\le N$ }


\State ${\bm s}_i^{\star}\leftarrow f_{\rm samp} ({\mathcal S})$

\State ${\mathcal S} \leftarrow [{\mathcal S}, \; {\bm s}_i^{\star}]$

\State $i = i+1$

\EndFor {\bf end}

\State \Return Library of OPF solutions $\mathcal S$

\end{algorithmic}}
\end{algorithm}

\begin{figure}[!ht]
\centering
\includegraphics[scale=0.65]{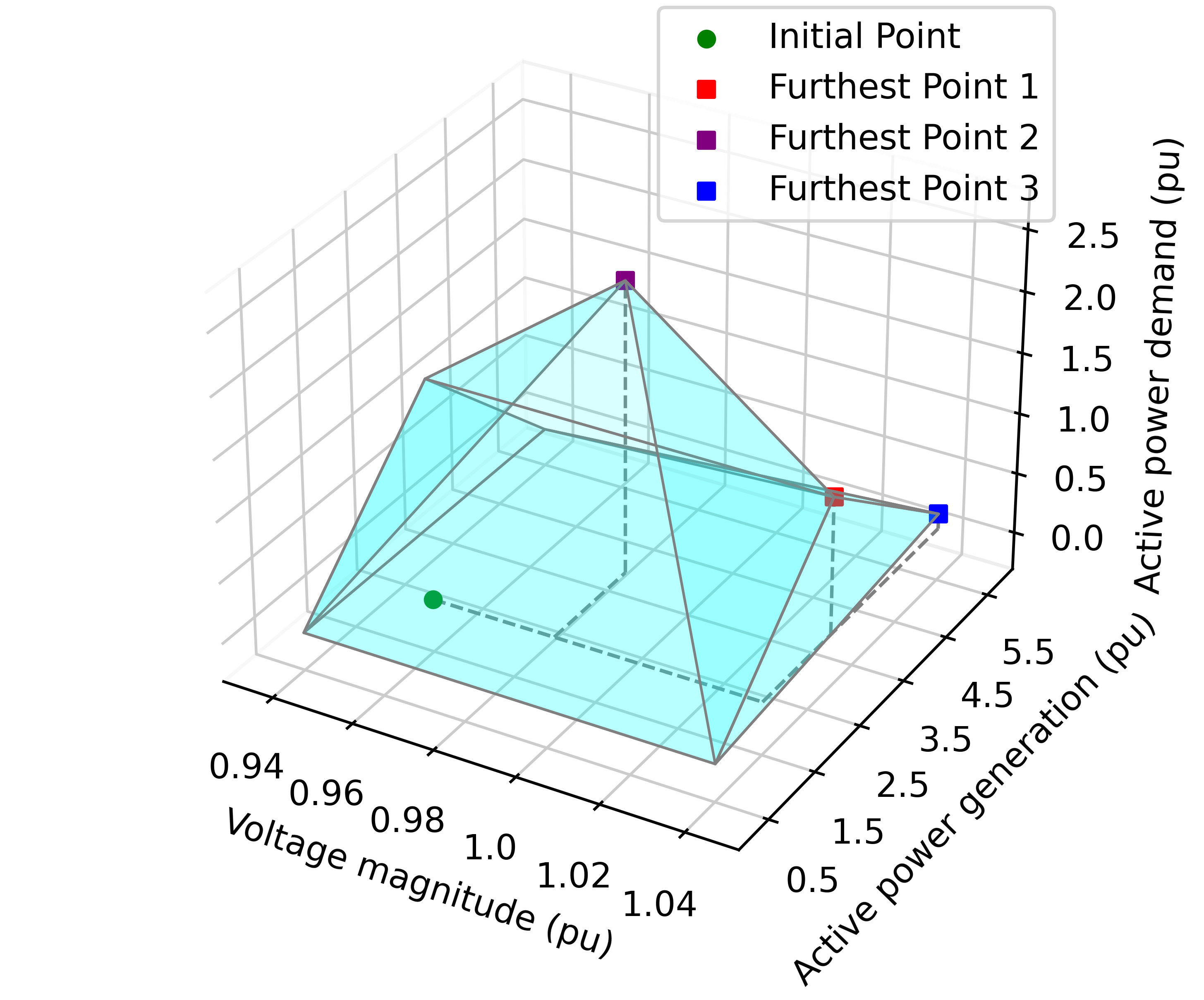}
\caption{Depicted is a 3D representation of the proposed bilevel optimization. The decision variables of the routine determine the axes: voltage magnitude, power generation, and demand. Within the feasible space (turquoise), three local optima are identified. 
These optima show how the OPF solution space is explored by maximizing distances between decision variables.} 
\label{distancerepresentation}
\end{figure}

Fig. \ref{distancerepresentation} illustrates the essence of the bilevel optimization in the case of three decision variables: voltage magnitude, active power demand and active power generation. During the first iteration, the optimizer looks for a point in the feasible space (i.e., within the turquoise surfaces) which is farthest from the initial green point; this is in Point 3 in blue. Once this point is located, it is added to the solution library $\mathcal S$ and the process continues (i.e., a new farthest point is found).
\subsection{Improving sampling routine scalability and solution quality}

Even with relaxation \eqref{eq: relax_eps}, the bilevel optimization problem \eqref{eq: first imp} poses a significant challenge for state-of-the-art solvers. Accordingly, the problem has a high dependency on initial conditions, and it also has scalability issues (both in the size of the power system considered, and in the number of power flow samples collected). In the following, we address both issues.

\subsubsection{Improving scalability via stochastic dataset and variable inclusion}
Summing over all previously found data points $\Gamma_\mathcal{S}$ in \eqref{upperlevelobj} becomes very expensive as the library of previously found points grows in size. To reduce this issue, each time we search for a new OPF solution, the library of previous OPF solutions seen by the objective function in \eqref{upperlevelobj} is limited to a random subset $\mathcal{S}_r \subset \mathcal{S}$, where $|\mathcal{S}_r| < |\mathcal{S}|$. Therefore, once the number of solutions in the library $\mathcal{S}$ surpasses some pre-defined threshold, each iteration will randomly pick a subset of solutions $\mathcal{S}_r$ from the full set $\mathcal{S}$ that will be used in the objective function. This strategy limits the amount of data seen by the objective function. 

Similarly, the algorithm can have scalability challenges in the context of large power systems with many buses. To ease this, we only sum over random subsets of buses ${\mathcal N}_r \subset {\mathcal N}$, generators ${\mathcal G}_r \subset {\mathcal G}$, and loads ${\mathcal D}_r \subset {\mathcal D}$ when building the objective function \eqref{upperlevelobj} at each sequential data collection iteration.

\subsubsection{Improving solution quality via stochastically filtered hot starts} We use a local, interior point method (Ipopt) to solve \eqref{eq: first imp}. To encourage the solver to find solutions which do not get stuck in local solution regions, random initialization was seen to be very helpful. Unfortunately, random initialization of \eqref{eq: first imp} also led to frequent solver divergence. To overcome this issue, we randomly initialized a friendlier version of \eqref{eq: first imp}, i.e., one which neglected both the stationarity and complementarity conditions in the KKT reformulation. The solution to this modified problem is a valid power flow solution (since primal feasibility is retained), but it isn't necessarily an \textit{optimal} power flow solution (since optimality conditions were dropped). The modified problem is given by
\begin{subequations}\label{eq: stage 1}
\begin{align}
\hat{f}_{{\rm samp}}(\mathcal{S})\,\triangleq\,{\max_{x,p_{d},q_{d}}}\quad & f_{{\rm dist}}(p_{G},p_{d},v,\mathcal{S})\\
{\rm s.t.}\quad & p_{d}^{{\rm min}}\leq p_{d}\leq p_{d}^{{\rm max}}\\
 & q_{d}^{{\rm min}}\leq q_{d}\leq q_{d}^{{\rm max}}\\
 & \eqref{eq: OPF_constraints}-\eqref{eq: inj}.
\end{align}
\end{subequations}
Therefore, on each data collection iteration, two problems are solved: \eqref{eq: stage 1}, then \eqref{eq: first imp}. Since the computationally challenging KKTs are dropped in the first problem, it is swiftly solved, and its solution is used to initialize the decision variables and the complementary slackness multipliers in the full problem.

Finally, to guarantee that sequentially obtained solutions to \eqref{eq: first imp} are indeed OPF solutions (some optimality error can be introduced via relaxation \eqref{eq: relax_eps}), a standard OPF problem is solved by fixing loads to the value found by the bilevel optimization. The solutions computed by this final OPF solver are the ones stored in the library of solutions $\mathcal S$\footnote{As new data points are sequentially found, we statistically analyze the differences between the relaxed bilevel solution and the final OPF solution.}. The final data collection routine, which we call the Routine for Aggregating Maximally Broadcasted Opfs (RAMBO) is stated in Alg.~\ref{algo:finaliterative}.
\begin{algorithm}
\caption{Final data collection procedure: \textbf{R}outine for \textbf{A}ggregating \textbf{M}aximally \textbf{B}roadcasted \textbf{O}pfs (\textbf{RAMBO}) }\label{algo:finaliterative}


{\small \textbf{Require:}
OPF model, number of data points $N$, $\epsilon$, subset sizes

\begin{algorithmic}[1]

\State Compute random feasible solution ${\bm s}^{\star}_1$ for \eqref{eqs: finalopf}

\State Set ${\mathcal S} = {\bm s}^{\star}_1$ and $i=2$

\For{$i\le N$ }

\State Sample a subset of data points ${\mathcal S}_r\subset{\mathcal S}$ 

\State Sample a subset of variables ${\mathcal N}_r$, ${\mathcal G}_r$, ${\mathcal D}_r$ $\subset$ ${\mathcal N}$, ${\mathcal G}$, ${\mathcal D}$

\State Initialize Ipopt by randomly perturbing primal variables

\State Solve \eqref{eq: stage 1}: $\Omega \leftarrow {\hat f}_{\rm samp}({\mathcal S}_r)$ with ${\mathcal N}_r$, ${\mathcal G}_r$, ${\mathcal D}_r$

\State Hot start Ipopt to solve $f_{\rm samp}$ with solution $\Omega$


\State Solve \eqref{eq: first imp}: ${\tilde{\bm s}}_i^{\star},P_d, Q_d\leftarrow f_{\rm samp} ({\mathcal S}_r)$ with ${\mathcal N}_r$, ${\mathcal G}_r$, ${\mathcal D}_r$


\State ${\bm s}_i^{\star} \leftarrow {\rm OPF}(P_d, Q_d)$

\State Collect verification statistics: $f_{\rm verify}(\{{\bm s}_i^{\star}, {\tilde{\bm s}}_i^{\star} \})$

\State ${\mathcal S} \leftarrow [{\mathcal S}, \; {\bm s}_i^{\star}]$

\State $i = i+1$

\EndFor {\bf end}

\State \Return Library of OPF solutions $\mathcal S$

\end{algorithmic}}
\end{algorithm}
\begin{figure}[!ht]
\centering
\includegraphics[width=\columnwidth]{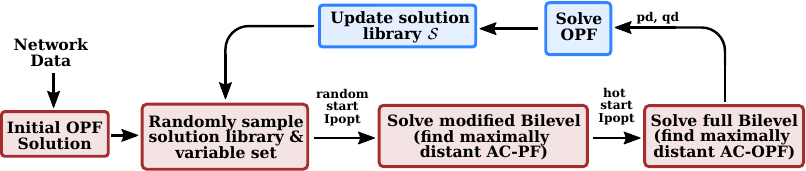}
\caption{Algorithmic flow of RAMBO (figure inspired by \cite{Jones:2022})}
\label{RAMBO}
\end{figure}


%% file: sec3.tex
Dataset quality is assessed primarily based on the number of active constraints the OPF solutions achieve; this is similar to the metric used by OPF-Learn~\cite{Jones:2022}. A constraint is considered active when the solution is very close to the limit imposed on the constraint, therefore having a nonzero dual variable. The RAMBO dataset is compared to a dataset generated using one of the most common approaches for collecting OPF data: random sampling of bounded loads.


\subsection{Binding constraints}
Four types of constraints were investigated in this paper: voltage limits, active and reactive power generation limits, and apparent power flow limits. The respective values from each solution of the datasets were compared with the aforementioned limits. If values were within 1\% of the limit value, the constraint was considered ``active".

\subsection{Random dataset creation}

Standard practice for AC-OPF dataset generation is to randomly sample within a small range centred on the nominal load of each bus. The sampling process typically uses a uniform distribution. Accordingly, the dataset used to benchmark RAMBO was created by sampling loads with a uniform distribution ranging from $-50\%$ to $+100\%$ of all nominal load values (i.e., $p^0_d$ could range from $p^0_d/2$ to $2 p^0_d$). Notably, we use the same load range when running RAMBO in the following experiments (i.e., when setting \eqref{secondcons}-\eqref{thirdcons}). Thus, both data sampling routines are exploring the exact same space.

%% file: sec4.tex
This section applies RAMBO (Alg. \ref{algo:finaliterative}) to the IEEE 30, 57 and 118-bus test systems from the PGLib library\cite{pglib}. The solver utilized for the optimization problems is Ipopt, an open-source interior point optimizer. Ipopt is a local solver suited for nonlinear programs and easily implemented in the Julia language and the PowerModels library. The results presented in this section consider $\epsilon=10^{-4}$, and the datasets contain $N=100$ samples each. Generally, a higher epsilon value reduces the computational time and accuracy of the program. At the same time, a more strict relaxation yields more accurate solutions but implies more complexity and infeasibility. In our analysis, $\epsilon=10^{-4}$ was seen to be an inflexion point with respect to speed, solution feasibility and accuracy. We also want to emphasize the importance of randomizing the objective function variables when using a local optimizer to solve this computationally intensive problem.

\subsection{Simulation results}
Here, we compare the performance of the datasets generated with the bilevel RAMBO approach against the uniformly sampled routine. The analysis has two primary focuses. The first is to evaluate which method accomplishes a better representation of the OPF solution space. This assessment is done by sampling problem decision variables. Subsequently, we can enrich the comparison by investigating which dataset captures the system's physical boundaries more effectively.

\begin{figure}[!ht]
\centering
\includegraphics[scale=0.55]{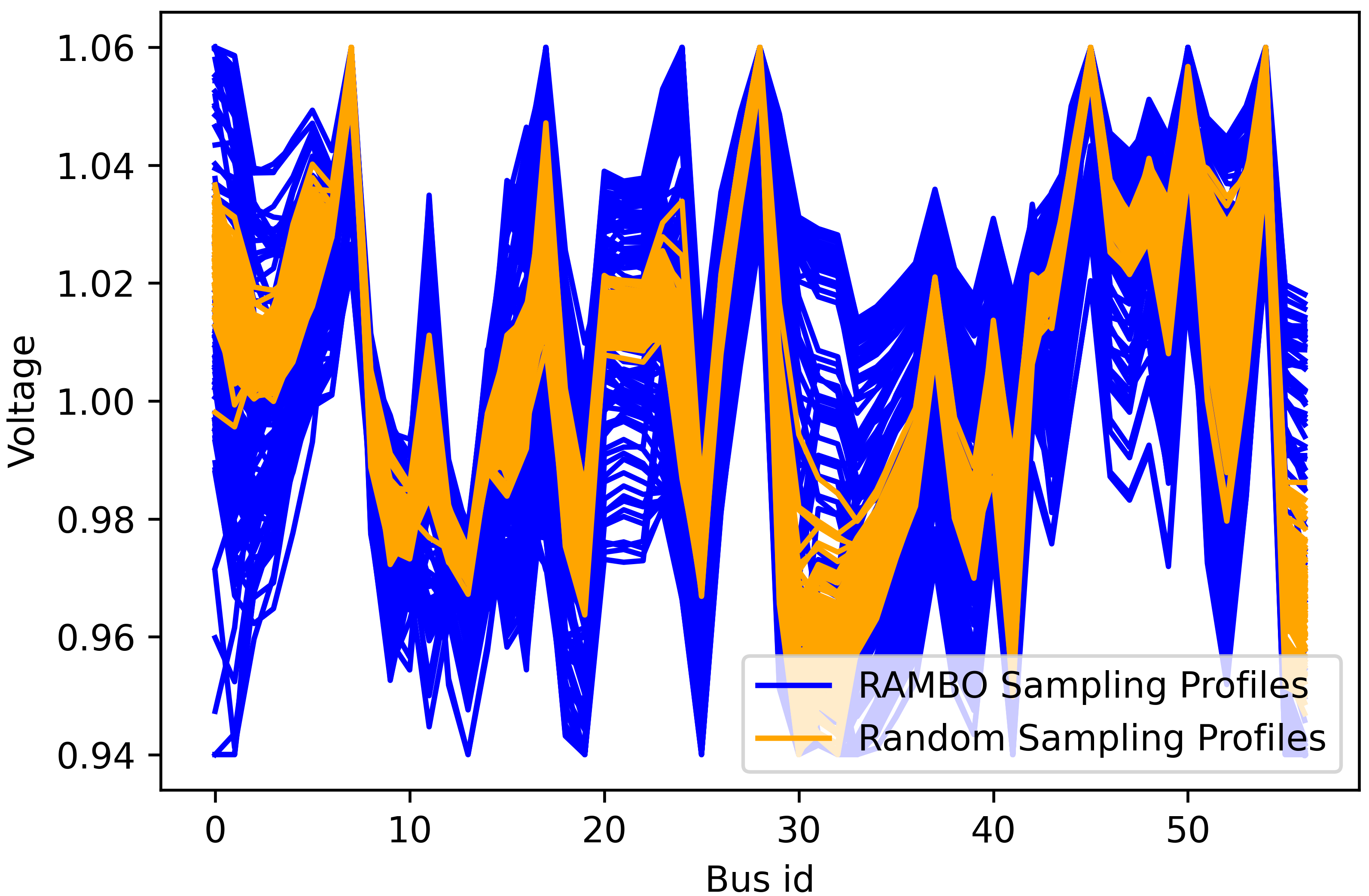}
\caption{Voltage magnitude feasible space sampled by the bilevel optimization (blue) and the uniform distribution (orange). The blue profiles cover a significantly broader region of the OPF solution space.}
\label{voltageprofile57}
\end{figure}

The following analysis reveals that a uniform sampling of the system's load, such as the standard routine applies, does not imply an accurate representation of the feasible space. Figure \ref{voltageprofile57} depicts this fact, where the voltage profiles of the proposed RAMBO routine significantly outperform the uniform sampling approach in terms of exploring more of the OPF solution space.\par
This fact is numerically demonstrated in Table \ref{tablediff}, which computes the average range of all decision variables across the system for all data. All decision variables (except the load $p_d$) have significantly larger ranges using the bilevel approach.

\begin{table}[!ht]
\begin{center}
\caption{Average difference between maximum and minimum values}\label{tablediff}
\begin{tabular}{c||c|c|c|c|c}
 \textbf{IEEE 57-bus}     &    \textbf{$v$}    &    \textbf{$p_G$}    &  \textbf{$q_G$}  & \textbf{$p_{ij}$} & \textbf{$p_d$}\\
 \hline
Uniform sampling    & 0.0239    & 1.973  & 0.572   & 0.319& 0.292\\
RAMBO    & 0.0667     & 4.535  &1.084  & 0.645& 0.298
\end{tabular}
\end{center}
\end{table}

Table \ref{table} compares the number of unique active constraints obtained with the uniform sampling approach vs with RAMBO. The flow column illustrates the number of line flows that are within 1\% of the branch rate limits. The voltage columns contain the number of bus voltages that are within 1\% of the maximum and minimum limits, respectively. Finally, the generation column describes the number of generators that operate within 1\% of their active and reactive power limits. 

RAMBO's performance in capturing the constraint limits in the solution space is significantly better compared to the uniform sampling method. Power flows, active and reactive generation, and bus voltages present a remarkable increase in unique binding constraints activated for RAMBO. The largest improvement is seen in the voltage lower limit set, where the standard routine struggles to capture low voltage profiles. Notably, in the 118-bus system, the random sampling approach didn't find a single active constraint set among minimum voltage limits. In contrast, RAMBO was effective in identifying as many as 48 unique active constraint sets.

\begin{table}[!ht]
\centering
\caption{Number of unique binding constraint sets}\label{table}
\begin{tabular}{c|c|c|c|c|c}
 \textbf{Method} & \textbf{System} & \textbf{Flow} & \textbf{Gen} & \textbf{Volt (min)} & \textbf{Volt (max)} \\
\hline
\multirow{3}{*}{Uniform} & 30-bus  & 1   & 4  & 0 & 3 \\
                         & 57-bus  & 0   & 7  & 2 & 5 \\
                         & 118-bus & 10  & 37 & \textbf{0} & 19 \\
\hline
\multirow{3}{*}{RAMBO}& 30-bus  & 1   & 7  & 4 & 5 \\
                      & 57-bus  & 0   & 10  & 12 & 8 \\
                      & 118-bus & 22  & 53 & \textbf{48} & 47 \\

\end{tabular}
\end{table}

Table \ref{tablesimilarity} highlights the error introduced by relaxation \eqref{eq: relax_eps}. Specifically, it depicts the average similarity between the solution of \eqref{eq: first imp} and the subsequent OPF solution (i.e., the difference between the right-most red box solution in Fig. \ref{RAMBO}, and the subsequent blue box solution). As expected, the similarity metric between the bilevel solutions compared to the OPF solutions is very close to 100\%. This reaffirms that the performance improvements achieved by relaxing the complementary slackness constraints are worth a slight difference from the exact optimal solution. However, because in the proposed method, the simple OPF problem computes the solutions library, the similarity metric is only used as confirmation that the load points are obtained through an accurate bilevel problem. 

\begin{table}[!ht]
\centering
\caption{Similarity between relaxed bilevel \& standard OPF solutions}
\label{tablesimilarity}
\begin{tabular}{c|c}
 \textbf{System} & \textbf{Similarity} \\
\hline
 30-bus  & 99.984\% \\
 57-bus  & 99.884\% \\
 118-bus & 99.976\% 
\end{tabular}
\end{table}

%% file: sec5.tex
This paper presents a new approach for generating high-quality OPF datasets, applicable to many future applications. We develop a method, RAMBO, that maximizes the distance between several OPF decision variables across sequential solutions to enhance the representation of the feasible space. We show that typical random load sampling approaches for OPF data collection do not accurately capture relevant state variables such as voltage or dispatch. These variables are included in the developed method to generate a more complete representation. Furthermore, the application to the PGLib networks shows a significant improvement in sampling critical data in the system boundaries. This characteristic makes it ideal for machine learning applications. Future work will add new decision variables, including dual variables associated with inequality constraints. Additionally, further effort will be directed towards scaling RAMBO to bigger power systems.